\documentclass[fleqn,10pt]{wlscirep}
\usepackage[utf8]{inputenc}
\usepackage[T1]{fontenc}
\usepackage{epsfig}
\usepackage{amssymb}
\usepackage{amsmath}
\usepackage{amsfonts}
\usepackage{graphicx}
\usepackage{mathrsfs}
\usepackage{multirow}
\usepackage{lipsum,appendix}
\usepackage{booktabs}
\usepackage{braket}

\newcommand{\fu}{\mathfrak{u}}

\newcommand{\fn}{{\mathfrak{n}}}

\newcommand{\fz}{\mathfrak{z}}

\newcommand{\fK}{\mathfrak{K}}

\newcommand{\bA}{\mathbf{A}}

\newcommand{\bC}{\mathbf{C}}

\newcommand{\bR}{\mathbf{R}}

\newcommand{\bT}{\mathbf{T}}

\newcommand{\cB}{\mathcal{B}}

\newcommand{\cD}{\mathcal{D}}
\newcommand{\cH}{\mathcal{H}}
\newcommand{\cE}{\mathcal{E}}

\newcommand{\cJ}{\mathcal{J}}

\newcommand{\be}{\begin{equation}}
\newcommand{\ee}{\end{equation}}
\newcommand{\bea}{\begin{eqnarray}}
\newcommand{\eea}{\end{eqnarray}}

\newcommand{\ed}{\end{document}}

\newcommand{\bi}{\begin{itemize}}
\newcommand{\ei}{\end{itemize}}

\newcommand{\bce}{\begin{center}}
\newcommand{\ece}{\end{center}}

\newcommand{\sE}{\mathscr{E}}
\newcommand{\sF}{\mathscr{F}}
\newcommand{\sG}{\mathscr{G}}

\title{Lasing with Topological Weyl Semimetal}

\author[1]{G\"{u}ne\c{s} Oktay}
\author[1,*]{Mustafa Sar{\i}saman}
\author[2]{Murat Tas}
\affil[1]{Department of Physics, Istanbul University, 34134, Vezneciler, Istanbul, Turkey}
\affil[2]{Department of Basic Sciences, Alt{\i}nba\c{s} University, 34217 Istanbul, Turkey}

\affil[*]{mustafa.sarisaman@istanbul.edu.tr}


\begin{abstract}
Lasing behavior of optically active planar topological Weyl semimetal (TWS) is investigated in 
view of the Kerr and Faraday rotations. Robust topological character of TWS is revealed by the 
presence of Weyl nodes and relevant surface conductivities. We focus our attention on the 
surfaces where no Fermi arcs are formed, and thus Maxwell equations contain topological terms. 
We explicitly demonstrate that two distinct lasing modes arise because of the presence of 
effective refractive indices which lead to the birefringence phenomena. Transfer matrix is 
constructed in such a way that reflection and transmission amplitudes involve $2\times2$ 
matrix-valued components describing the bimodal character of the TWS laser. We provide 
associated parameters of the topological laser system yielding the optimal impacts. We reveal 
that gain values corresponding to the lasing threshold display a quantized behavior, which 
occurs due to topological character of the system. Our proposal is supported by the 
corresponding graphical demonstrations. Our observations and predictions suggest a concrete 
way of forming TWS laser and coherent perfect absorber; and are awaited to be confirmed by an 
experimental realization based on our computations.
\end{abstract}
\begin{document}

\flushbottom
\maketitle
%
%
\thispagestyle{empty}

\section*{Introduction}

Introducing topological ideas into the material science has led to the emergence of a new and 
impressive field of physics, topological materials~\cite{topmat1, topmat1-1}. Especially, 
their interplay with electromagnetic waves offers remarkable applications in aspects of 
photonics, yielding a recent field of physics, topological photonics~\cite{topphot1, topphot1-1}. 
Recently, numerous active studies are witnessed in this field extending to various intriguing 
directions, which also help various aspects of non-Hermitian quantum mechanics to be understood 
in view of topological perspective. In this respect, recent study of constructing a theoretical 
modeling of topological insulator lasers and their accompanying experimental realization 
attracted attention to this field, which has served to alter current understanding of the 
relation between disorder and lasing, and has opened exciting possibilities at the interface of 
topological physics and laser science, such as topologically protected transport in systems 
with gain.~\cite{topinsullaser, topinsullaser-1}. Motivated by these recent findings, we build 
up a theoretical modeling of a topological Weyl semimetal (TWS) laser by employing the transfer 
matrix approach in this paper. We reveal that topological protection of the surface states 
gives rise to bimodal topological laser which requires quantized gain values for the 
realization of the lasing threshold condition.

The TWS, one of leading candidates of topological materials, has attracted much attention in 
recent years for its novel gapless band structure in the bulk and exotic Fermi arcs on the 
surface states. Theoretical predictions of this exclusive phase of topological materials has 
been made since 2011~\cite{theorpredic, theorpredic-1, theorpredic-2, theorpredic-3, 
theorpredic-4}, and the first experimental verification of this phase was reported in TaAs 
material in 2015~\cite{experpredic, experpredic-1}. Recently, great interest in the TWS phase 
has arisen due to the possibility of realization of this phase in certain condensed matter 
systems. There are various predictions of solid materials as potential TWS 
candidates~\cite{potentialTWS, potentialTWS-1, potentialTWS-2}, although their energy 
structures and topological properties are not perfectly consistent with an ideal TWS. In fact, 
an ideal TWS proposed in~\cite{potentialTWS} is rare in nature, and makes the analysis of the 
corresponding properties of TWS, i.e., spin texture of the boundary states, the flat bands, 
and the transport properties, quite difficult~\cite{idealTWS2, idealTWS2-1}. Their interaction 
with electromagnetic waves is thus quite intriguing, and gives rise to substantial 
applications of them, especially in developing new photonics devices. In this study, we 
fulfill a comprehensive treatment of electromagnetic scattering of these attractive materials 
just to unveil their lasing features and corresponding CPA applications.

Optical properties of TWS materials can be specified via their optical conductivities. For 
this purpose, the bulk and surface conductivity tensors taking into account all possible 
combinations of the optical transitions involving bulk and surface electron states can be 
employed. Thus, one can show how the information about electronic structure of TWS materials, 
such as the position and separation of Weyl nodes, Fermi energy, and Fermi arc surface states, 
can be extracted unambiguously from measurements of the dispersion, transmission, reflection, 
and polarization of electromagnetic waves\cite{Qchen, Moore, Grassano}. As for the optical 
behaviors of TWS in a laser pulse, the femtosecond dynamics of electrons in such materials 
is highly irreversible, i.e., the residual electron population after the pulse is comparable 
to the conduction-band population during the pulse. Irreversibility of ultrafast electron 
dynamics is determined not only by the band gap of the material, but also by the profile and 
magnitude of the interband dipole elements. In the case of circularly-polarized pulse, 
response of the system to the linear probe pulse is highly dependent on the intrinsic 
chirality of the Weyl nodes. The magnitude of the charge transfer strongly depends on the 
direction of polarization\cite{Neubauer, Nematollahi, Nematollahi2}.

An optically active laser system can be obtained by means of the existence of spectral 
singularities~\cite{naimark, naimark-1, p123}. Hence, we mount a homogeneous gain environment 
in a planar TWS slab such that TWS properties are preserved in a way that corresponding Weyl 
nodes are not affected, and Fermi arcs are present on its surfaces. Once electromagnetic 
waves are incident on a surface which does not contain a Fermi arc, polarization directions 
of the reflected and transmitted waves change due to the Kerr and Faraday rotations on the 
surfaces. This leads to the modified Maxwell equations involving topological terms in the 
source parts. Solution of these equations reveals coupled Helmholtz equations whose solutions 
yield a $4\times4$ transfer matrix. We show that the form of the transfer matrix provides a 
basis for the bimodal topological laser. This is characteristics of a TWS laser system such 
that purely outgoing waves cause rotated output intensity due to the Kerr and Faraday 
rotations. A coherent perfect absorber (CPA) using TWS can be thus obtained by adjusting 
appropriate polarizations given by the Faraday rotations together with exact phase and 
amplitude modulations~\cite{CPA, lastpaper, CPA-8, CPA-9}. This happens by virtue of the 
fact that CPA has the time reversal symmetry of a laser.

\begin{figure}
    \begin{center}
    \includegraphics[scale=.50]{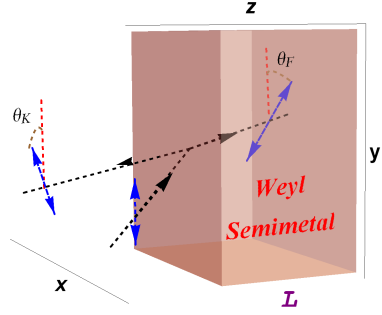}
    \caption{(Color online) The TE mode configuration for the interaction of a electromagnetic 
wave with the Weyl semimetal slab. The wave is emitted on the slab by an angle $\theta$ which 
is measured from the normal to the surface, and direction of the polarization is rotated by 
an angle of $\theta_F$ and $\theta_K$ inside and outside of the slab respectively.}
    \label{fig1}
    \end{center}
    \end{figure}

Although original problem is one dimensional, Kerr and Faraday rotations turn the problem 
into two-dimensional plane, see Fig.~\ref{fig1}. Therefore, reflection and transmission 
amplitudes are expressed by means of $2\times2$ matrices. Dimension of these matrices 
decides bimodal structure of a TWS laser system. The spectral singularities for each mode 
are established by the divergent characteristics of reflection and transmission matrices. 
Hence, we obtain uni- or bimodal spectral singularity conditions with/without dispersion 
effects. We look for practical ways to improve the efficiency of the topological laser and 
CPA systems through the supplemented material features. It is noted that our formalism is 
satisfied by all TWS type materials.

We find complete solutions, schematically demonstrate their behaviors and deduce the effects 
of various parameter choices yielding laser conditions in the framework that we develop. We 
reveal the optimal control of parameters in a TWS uni- or bimodal slab laser, which include 
gain coefficient, incidence angle, slab thickness, and Weyl node separation. These optimal 
parameters give rise to a desired outcome of achieving rotated outgoing waves in the uni- or 
bimodal TWS laser system. We also present the way to find exact conditions causing the 
achievement of TWS CPA with equal amplitude and phase values of ingoing waves which are 
obtained by adjusting correct Faraday rotation angles. Our method, and thus results guide 
possible experimental studies in this direction for all proposed TWS slab materials of 
practical concern.

\section*{TE Mode Solution and Transfer Matrix}
\label{S2}

Consider a linear homogeneous and optically active gain slab system which is made up of a TWS 
material whose Weyl nodes are aligned along the $z$-axis as depicted in~Fig.~\ref{fig1}. The 
slab is designed in such a way that it has a thickness $L$ and a complex refractive index 
$\fn$ which is uniform between the end-faces of the slab in the region $0<z<L$. Interaction 
of this TWS optical slab with the electromagnetic waves requires an elaborate analysis of 
the properties of TWS in view of topological and magnetoelectric optical effects. Topological 
features are crucial based on the alignment of the Weyl nodes which specify locations of the 
Fermi arcs on the faces of the slab. In our set-up, Fermi arcs appear on the side faces of 
the slab along the $z$-direction since the Weyl nodes are oriented in $z$-axis as shown in 
Fig.~2. In the light of this optical setup, Maxwell equations turn out to include 
topological terms (see Appendix A for the derivation of the topological terms) which are 
shown to take the form of

    \begin{figure}
    \begin{center}
    \includegraphics[scale=.50]{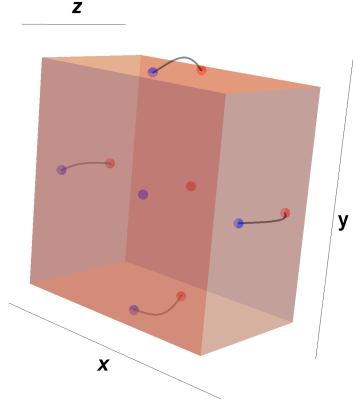}
    \caption{(Color online) Schematic representation of locations of the Weyl nodes along 
the $k_z$-axis in bulk Brillouin zone, which are the sources (green dot) and sinks (pink dot) 
of the Berry curvature. Fermi arcs corresponding to the Weyl nodes are formed on the relevant 
surfaces.}
    \label{fig2}
    \end{center}
    \end{figure}

    \begin{align}
    &\vec{\nabla}\cdot\vec{\cD} = \rho (z) + \beta\,\vec{b}\cdot\vec{\cB}, &&
    \vec{\nabla}\cdot\vec{\cB} = 0,
    \label{equation1}\\
    &\vec{\nabla} \times \vec{\cH}-\partial_t \vec{\cD}=\vec{\cJ}(z) -\beta\,\vec{b}\times\vec{\cE}, &&
    \partial_t \vec{\cB}+\vec{\nabla} \times \vec{\cE}=\vec 0.
        \label{equation2}
    \end{align}
 Here $\beta := 2\alpha / \pi Z_0$ is a constant, $\alpha:=e^2 / 4\pi\varepsilon_0\hbar c$ is 
the fine structure constant, $Z_0:=\sqrt{\mu_0/\varepsilon_0}$ is the vacuum impedance, $e$ is 
the charge of an electron, and $c:= 1/\sqrt{\varepsilon_0 \mu_0}$ is the speed of light in 
vacuum. The vector notation $\vec{b}$ specifies the distance between two Weyl nodes which are 
aligned in the $z$-direction and is given explicitly by $\vec{b} (z)= b (z) \hat{e}_z$, and 
$b (z) = b\,\fu(z)\, \fu(L-z)$, where $\fu(z)$ is the Heaviside step function defined as
   \be
   \fu (z):= \left\{\begin{array}{cc}
    0 & {\rm for}~~~~  z<0\\
    1 & {\rm for}~~~~  z>0
    \end{array}\right. \notag
   \ee
 Notice that $\vec{\cE}$ and $\vec{\cB}$ represent the electric and magnetic fields, 
respectively, and are expressed by $\vec{\cD}$ and $\vec{\cH}$ fields via the following 
constitutive relations
    \begin{align*}
    \vec{\cD} := \tilde{\varepsilon}\, \vec{\cE}, &&\vec{\cB}:=\tilde{\mu}\vec{\cH},
    \end{align*}
where $\tilde{\varepsilon}$ and $\tilde{\mu}$ are, respectively, permittivity and 
permeability of the relevant environment in which the electromagnetic wave propagates. 
These are described by $\tilde{\varepsilon} := \varepsilon_0 \varepsilon$ and 
$\tilde{\mu} := \mu_0 \mu$, respectively, in terms of the vacuum permittivity and 
permeability, where we defined
    \begin{align}
    \varepsilon (z)&:= \left\{\begin{array}{cc}
    \varepsilon_b + \frac{i \sigma_{yy}}{\varepsilon_0 \omega} & {\rm for}~~~~  z \in [0, L],\\
    1 & {\rm otherwise}
    \end{array}\right. \label{e1}\\
    \mu (z)&:= \left\{\begin{array}{cc}
    1 + \chi_m & {\rm for}~~~~ z \in [0, L],\\
    1 & {\rm otherwise}
    \end{array}\right.
    \label{e2}
    \end{align}
$\varepsilon_b$ is the bound charge contribution, $\chi_m$ is the magnetic susceptibility 
of the TWS provided that it exhibits a magnetic characteristics. In this study, we assume 
that TWS is endowed with a rather weak magnetism such that we may ignore it~\footnote{For 
the sake of full discussion, we keep its presence till the end.}. Notice that 
$\fn^2 := \varepsilon \mu$ within the slab according to the expressions (\ref{e1}) and 
(\ref{e2}), where $\fn$ corresponds to the complex-valued refractive index of TWS, and is 
described by real and imaginary parts as follows
   \be
   \fn := \eta + i\kappa \label{refractive}
   \ee
We observe that free charges and currents in our optical setup appear only on surfaces of 
the TWS slab on which the incident wave emerges and no Fermi arc is present. This is 
because of the topological character of TWS which leads to form conductivities only on 
the surfaces, whereas it behaves as a semimetal inside the TWS medium. Therefore, surface 
charge density $\rho^s (z)$ and conductivity $\sigma^s(z)$ can be written as
    \begin{align*}
    \rho^s (z) := \rho^{(1)} \delta(z) + \rho^{(2)} \delta(z-L),\notag\\
    \sigma^s (z) := \sigma^{(1)} \delta(z) + \sigma^{(2)} \delta(z-L),\notag
    \end{align*}
where $\rho^{(j)}$ and $\sigma^{(j)}$ are, respectively, the free charge and conductivity on 
the $j$-th layer, with $j = 1, 2$.  Notice that $\rho^s (z)$ and $\sigma^s (z)$ are 
associated to each other by the continuity equation
    \be
    \vec{\nabla}\cdot\vec{\cJ}^s + \partial_t \rho^s (z) = 0
    \ee
for the electric current density given by $\vec{\cJ}^s := \sigma^s(z) \vec{\cE}$. Surface 
conductivity $\sigma^s$ has a tensorial expression, and thus we can demonstrate the surface 
current $\vec{\cJ}^s$ as $\vec{\cJ}^s_{\alpha} = \sigma^s_{\alpha\beta} \vec{E}_{\beta}$. 
In our case,
   \begin{align}
   \sigma^s_{yy} &=\frac{e^2 k_c}{3 \pi h \hat{\omega}_c} \left\{1 -i \left[\hat{\omega}_c^2+\ln\left|1-\hat{\omega}_c^2\right|\right]\right\}   \label{conductivity1}\\
   \sigma^s_{yx} &=\frac{e^2 b}{\pi h} + \frac{\alpha c}{3 \pi v_F} \ln\left|1-\hat{\omega}_c^2\right|   \label{conductivity2}
   \end{align}
where $\hat{\omega}_c := 2\omega_c/\omega$,  $\omega_c := v_F k_c$, $v_F$ is the Fermi 
velocity, $k_c$ is the momentum cut-off and $k \leq k_c$, see Appendix for the details of 
the derivations of (\ref{conductivity1}) and (\ref{conductivity2}). We emphasize that the 
conductivity $\sigma^s_{yx}$ is responsible for the Kerr and Faraday rotations inside and 
outside the TWS. Consider time harmonic electromagnetic waves~\footnote{Time harmonic 
electric field corresponds to $\vec{\cE} (\vec{r}, t):= \vec E(\vec{r})\,e^{-i\omega t}$, 
similarly for $\vec{\cB} (\vec{r}, t)$, $\vec{\cD} (\vec{r}, t)$ and $\vec{\cH} (\vec{r}, t)$ 
fields.} having TE mode solutions corresponding to our geometrical design such that the wave 
can be regarded as obliquely incident in the form, see Fig.~\ref{fig1}
    \begin{align}
    \vec E(\vec{r})=\sE (z)e^{ik_{x}x}\hat e_y,
    \label{ez1}
    \end{align}
In this notation, $\hat e_j$ denotes the unit vector along the $j$-axis, with 
$j = x, y\,\textrm{and}\,z$, and $k_j$ is the component of wavevector $\vec{k}$ in the 
$j$-direction, where $\vec{k}$ is given by
    \be
    \vec{k} = k_x\,\hat{e}_x + k_z\,\hat{e}_z \notag
    \ee
such that $k_x = k\,\sin\theta$, $k_z = k\,\cos\theta$, and $\theta \in [-90^{\circ},90^{\circ}]$ 
represents the incident angle. We notice that polarization direction of the emitted wave 
will be twisted once it is reflected and refracted at the interface of TWS. Reflected waves 
give rise to the Kerr rotation while the refracted waves to the Faraday rotation within the 
slab, see Fig.~\ref{fig1}.

\begin{figure*}
    \begin{center}
    \includegraphics[scale=.40]{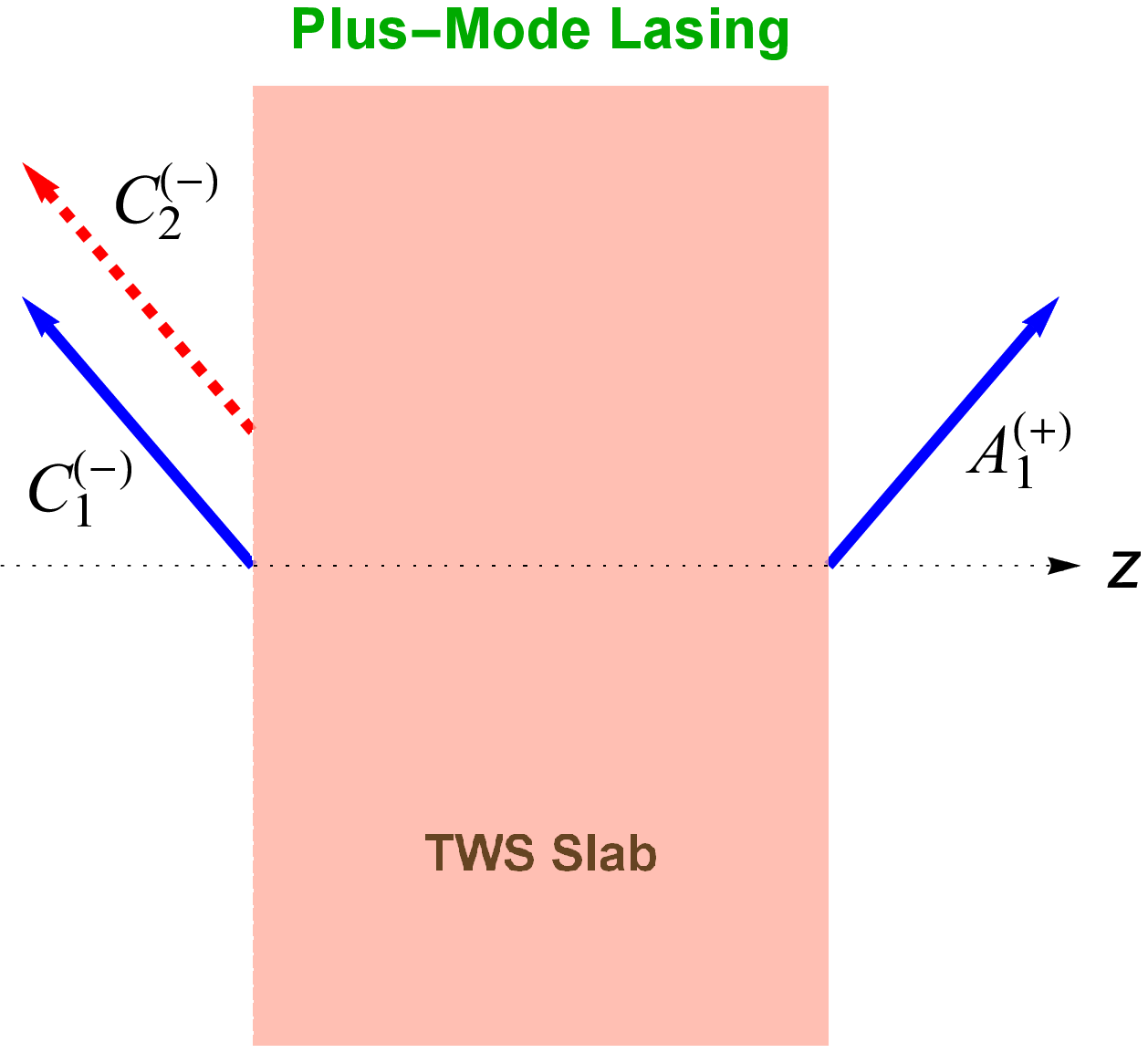}~~
    \includegraphics[scale=.40]{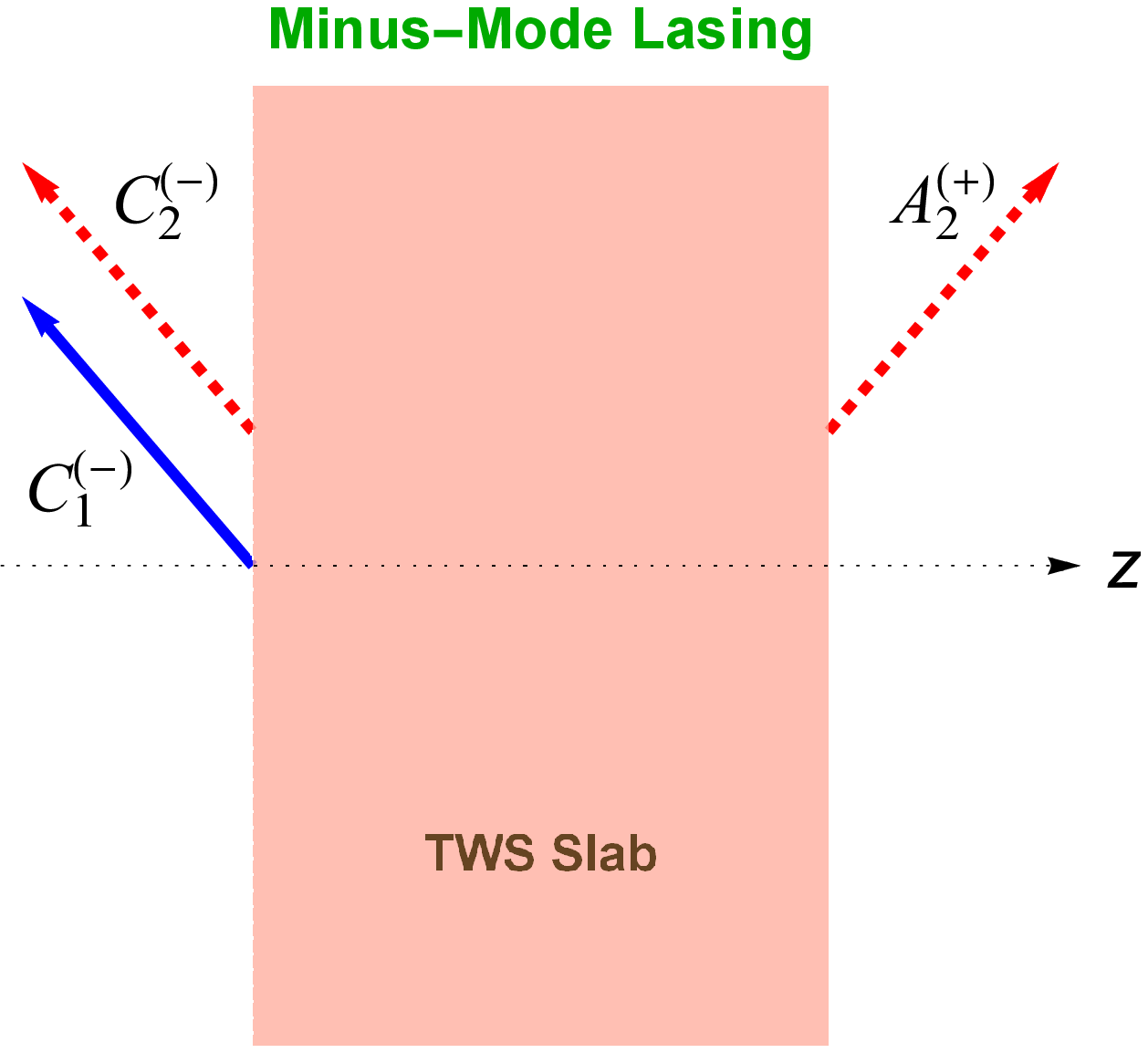}~~
    \includegraphics[scale=.40]{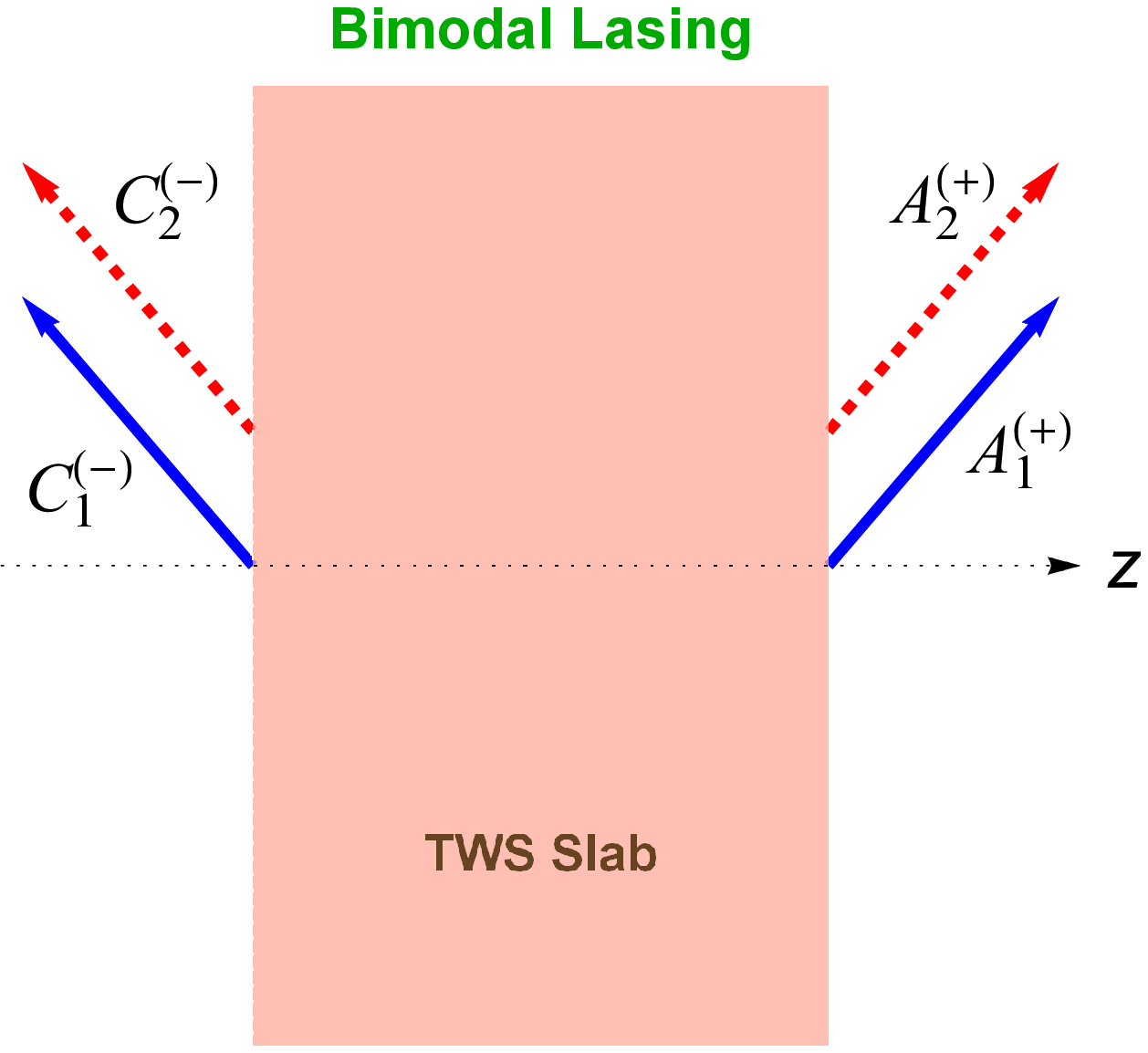}
    \caption{(Color online) Configurations of uni- and bimodal lasing. Blue solid arrows 
correspond to lasing due to $\tilde{\fn}_{+}$, and red dashed arrows correspond to lasing 
due to $\tilde{\fn}_{-}$.}
    \label{figplusminuslasing}
    \end{center}
    \end{figure*}

The Maxwell equations in (\ref{equation1}) and (\ref{equation2}) can be manipulated to give 
3-dimensional Helmholtz equation associated with the TE mode states, and corresponding 
magnetic field $\vec{H}$ as follows
   \begin{align}
   &\left[\nabla^{2} +k^2\varepsilon(z)\mu(z)\right]\vec{E} -i\beta k Z_0 \vec{b}\times\vec{E} = 0,\label{helmholtz1}\\ &\vec{H} = -\frac{i}{k Z_{0}\mu(z)} \vec{\nabla} \times \vec{E},\label{helmholtz2}
   \end{align}
TE mode solution~\footnote{We obtain the TE mode solution of the Kerr and Faraday rotated 
optical system by specifying $\vec{E} = E_x \hat{e}_x + E_y \hat{e}_y$, where 
$E_x (\vec{r}) = \sE_x (z)e^{ik_{x}x}$ and $E_y (\vec{r}) = \sE_y (z)e^{ik_{x}x}$.} of 
(\ref{helmholtz1}) in view of the Kerr and Faraday rotations can be established by means of 
the following one-dimensional coupled Helmholtz equations
  \begin{align}
  \sE''_x + k_z^2 \fz(z) \sE_x +i\beta k Z_0 b\,\sE_y = 0 \label{helm1}\\
  \sE''_y + k_z^2 \fz(z) \sE_y -i\beta k Z_0 b\,\sE_x = 0 \label{helm2}
  \end{align}
where a prime denotes derivative with respect to $z$. The piecewise constant function 
$\fz(z)$ is specified by
    \begin{align}
    \fz(z)&:=\left\{\begin{array}{cc}
    \tilde{\fn}^{2} & {\rm for}~~~~ z \in [0, L],\\
    1 & {\rm otherwise}
    \end{array}\right.\notag\\
    \tilde{\fn}&:= \sec\theta\sqrt{\fn^2 -\sin^2\theta} \notag
    \end{align}
We next introduce the following scaled variables for the convenience of subsequent expressions
    \begin{align}
    &\mathbf{x}:=\frac{x}{L}, &&\mathbf{z}:=\frac{z}{L},  &&\fK:=Lk_z=kL\cos\theta . \label{scaled-var}
    \end{align}
Solutions of (\ref{helm1}) and (\ref{helm2}) are attained once we split them in uncoupled 
modes as follows
    \be
    \psi_{\pm} (\mathbf{z}) := \sE_x (L \mathbf{z}) \pm i \sE_y (L \mathbf{z}) \notag
    \ee
where $\psi_{\pm}$ are the solutions of Schr\"{o}dinger equations
    \be
    -\psi''_{\pm} +  v_{\pm}(\mathbf{z}) \psi_{\pm} = \fK^2 \psi_{\pm} \label{schro}
    \ee
for the potentials given by $v_{\pm}(\mathbf{z}) = \fK^2 \fz_{\pm} (\mathbf{z})$. Here 
$\fz_{\pm} (\mathbf{z})$ is defined as
    \be
    \fz_{\pm} (\mathbf{z}) := \fz (\mathbf{z}) \pm \frac{2\alpha L b(\mathbf{z})}{\pi \fK \cos\theta} \label{birefrin}
    \ee
Notice that refractive indices $\tilde{\fn}_{\pm} = \sqrt{\tilde{\fn}^2 \pm 2\alpha b L /\pi \fK \cos\theta}$ within the TWS slab lead to the birefrigence effect because of the presence of uncoupled modes in (\ref{birefrin}). In view of (\ref{helmholtz2}), (\ref{schro}) and (\ref{birefrin}), one finds components of the electric field $\vec{E}$ and magnetic field $\vec{H}$ as in Table~\ref{table01},
\begin{table*}[ht]
\centering
{%
\begin{tabular}{@{\extracolsep{4pt}}llllcccc}
\toprule
{} & \multicolumn{1}{c}{Components of $\vec{E}$-field} & {} & \multicolumn{1}{c}{Components of $\vec{H}$-field} & {} \\
 \cline{2-2}
 \cline{3-5}
 \cline{6-8}
   \hline
  & $E_x = \frac{(\sF_+ + \sG_+)}{2}\,e^{i\fK \mathbf{x}\tan\theta}$ & {} & $H_x = \frac{i\cos\theta}{2 Z_0}\left[\sqrt{\fz_{+}}\sF_- - \sqrt{\fz_{-}}\sG_-\right]e^{i\fK \mathbf{x}\tan\theta}$ & {} \\
  & $E_y = \frac{-i(\sF_+ - \sG_+)}{2}\,e^{i\fK \mathbf{x}\tan\theta}$ & {} & $H_y = \frac{\cos\theta}{2 Z_0 \mu}\left[\sqrt{\fz_{+}}\sF_- + \sqrt{\fz_{-}}\sG_-\right]e^{i\fK \mathbf{x}\tan\theta}$ & {} \\
  & $E_z = 0$ & {} & $H_z = -\frac{i\sin\theta}{2 Z_0}\left[\sF_+ - \sG_+\right]e^{i\fK \mathbf{x}\tan\theta}$ & {} \\
 \hline
\end{tabular}%
}
\caption{Components of $\vec{E}$ and $\vec{H}$ fields existing inside and outside the TWS slab.}\label{table01}
\end{table*}
where the quantities $\sF_{\pm}$ and $\sG_{\pm}$ are defined in different regions of optical 
TWS slab system by
    \begin{flalign}
    \sF_{\pm}&:=\left\{\begin{array}{ccc}
    A_1^{(-)}\,e^{i\fK\mathbf{z}} \pm C_1^{(-)}\,e^{-i\fK\mathbf{z}} & {\rm for}~~ \mathbf{z} < 0,\\
    B_1^{(+)}\,e^{i\fK_+\mathbf{z}} \pm B_2^{(+)}\,e^{-i\fK_+\mathbf{z}} & {\rm for}~~ 0 < \mathbf{z} < 1 ,\\
    A_1^{(+)}\,e^{i\fK\mathbf{z}} \pm C_1^{(+)}\,e^{-i\fK\mathbf{z}} & {\rm for}~~ \mathbf{z} > 1.
    \end{array}\right. ~~~~~~
    \sG_{\pm} :=\left\{\begin{array}{ccc}
    A_2^{(-)}\,e^{i\fK\mathbf{z}} \pm C_2^{(-)}\,e^{-i\fK\mathbf{z}} & {\rm for}~~ \mathbf{z} < 0,\\
    B_1^{(-)}\,e^{i\fK_-\mathbf{z}} \pm B_2^{(-)}\,e^{-i\fK_-\mathbf{z}} & {\rm for}~~ 0 < \mathbf{z} < 1 ,\\
    A_2^{(+)}\,e^{i\fK\mathbf{z}} \pm C_2^{(+)}\,e^{-i\fK\mathbf{z}} & {\rm for}~~ \mathbf{z} > 1.
    \end{array}\right.\notag
    \end{flalign}
Here we introduced the quantity $\fK_j$ as follows
    \be
    \fK_j := \fK \tilde{\fn}_j. \label{ktildej}
    \ee
The complex coefficients $A_j^{(\pm)}, B_j^{(\pm)}$ and $C_j^{(\pm)}$ are possibly 
$\fK$-dependent, and related to each other by means of standard boundary conditions. 
Appropriate boundary conditions in our configuration of the optical system are described by 
the fact that tangential components of $\vec{E}$-fields and normal components of 
$\vec{B}$-fields are continuous across the interfaces while tangential components of 
$\vec{H}$-fields are discontinuous by an amount equal to the surface current density 
$\vec{\cJ}$ which is expressed by the conductivities on the corresponding surfaces. See 
Appendix for the associated boundary conditions. Hence, the transfer matrix following from 
the boundary conditions are obtained as follows
   \be
   \left(
           \begin{array}{c}
             \bA^{(+)} \\
             \bC^{(+)} \\
           \end{array}
   \right) = \mathbb{M} (\fK)\left(
           \begin{array}{c}
             \bA^{(-)} \\
             \bC^{(-)} \\
           \end{array}
   \right). \notag
   \ee
Here $\bA^{(\pm)}$ and $\bC^{(\pm)}$ are the column matrices which represent the coefficients 
of right and left moving waves outside the TWS slab, and are given by
\begin{align}
&\bA^{(\pm)} = \left(
                       \begin{array}{c}
                         A_1^{(\pm)} \\
                         A_2^{(\pm)} \\
                       \end{array}
                     \right), ~~~~~\bC^{(\pm)} = \left(
                       \begin{array}{c}
                         C_1^{(\pm)} \\
                         C_2^{(\pm)} \\
                       \end{array}
                     \right),\notag
\end{align}
and $\mathbb{M} (\fK)$ is the $4\times4$ transfer matrix~\cite{prl-2009} which is expressed by
\be
\mathbb{M} (\fK) = \left(
                       \begin{array}{cc}
                         \bT^{l}- \bR^{l}\bR^{r}\bT^{-r}& \bR^{r}\bT^{-r} \\
                         -\bR^{l}\bT^{-r} & \bT^{-r} \\
                       \end{array}
                     \right)\label{transmatr}
\ee
Once we impose the reciprocity principle, we understand that $\bT^{l} = \bT^{r} = \bT$. 
Notice that the transfer matrix (\ref{transmatr}) gives rise to deduce all information about 
the lasing properties of TWS optical configuration since it contains valuable information 
about reflection and transmission amplitudes. Lasing threshold condition is then given by 
the spectral singularity expression which is given by the requirement $\mathbb{M}_{22} = 0$ 
corresponding to the real $\fK$ values. Thus, transmission and right/left reflection 
amplitudes are divergent at spectral singularity points. In our TWS configuration, 
$\mathbb{M}_{22}$ is obtained as follows
\be
\mathbb{M}_{22} = \bT^{-1} = -\left(
                               \begin{array}{cc}
                                 \zeta_-(\nu_+; \nu_-) & \gamma_-(\nu_+; \nu_-) \\
                                 \gamma_-(\nu_-; \nu_+) & \zeta_-(\nu_-; \nu_+) \\
                               \end{array}
                             \right)\,e^{i\fK},\label{M22}
\ee
where $\nu_{j} := (\sigma_j, \tilde{\fn}_j)$, and $\zeta_{j}$, $\gamma_{j}: \mathbb{C}\rightarrow\mathbb{C}$ are continuous functions and specified by
\begin{align*}
\zeta_j(\nu_+; \nu_-) := &\frac{1}{2}\Bigl\{\left(\sigma_+-\sigma_- + 2j\right)\cos\fK_+ -\frac{i\mu\sigma_+^2}{\tilde{\fn}_-}\sin\fK_-
+ \frac{i\left[\tilde{\fn}_+^2 -\mu^2 \left(\sigma_+ + j\right)\left(\sigma_- - j\right)\right]}{\mu\tilde{\fn}_+}\sin\fK_+\Bigr\},\notag\\
\gamma_j(\nu_+; \nu_-) := &\frac{1}{2}\Bigl\{\sigma_+ \cos\fK_- + \frac{i\mu \sigma_+\left(\sigma_--1\right)}{\tilde{\fn}_-}\sin\fK_-
-\sigma_- \cos\fK_+ + \frac{i\mu \sigma_-\left(\sigma_--j\right)}{\tilde{\fn}_+}\sin\fK_+\Bigr\},\notag
\end{align*}
with the appropriate identification of $\sigma_j$
\begin{align}
   \sigma_j := \frac{\mu_0 \omega L}{2\fK} \left(\sigma_{yy} + i j \sigma_{yx}\right) = \frac{Z_0}{2\cos\theta}\left(\sigma_{yy} + i j \sigma_{yx}\right) \label{sigmapl}
\end{align}
where $j = +~\textrm{or}~-$ is implied. Thus, lasing threshold condition of TWS is obtained 
by means of the spectral singularities which are found by the real values of $\fK$ such that 
matrix components of $\mathbb{M}_{22}$ in Eq.~\ref{M22} are set to zero. In view of this fact, 
we realize that each of the plus ($``+"$) and minus ($``-"$) modes corresponding to the 
plus/minus refractive indices may lead to form distinctive lasing conditions. This two-mode 
lasing arises due to the coupling behavior of solutions appearing in the TWS configurations, 
which is seen by virtue of the Kerr and Faraday rotations. Thus, plus/minus mode lasing is 
attained by the condition
\be
\zeta_-(\nu_{\pm}; \nu_{\mp}) = \gamma_-(\nu_{\pm}; \nu_{\mp}) = 0.\label{twomodelasing}
\ee
This in turn implies the conditions $A_1^{(-)} = A_2^{(-)} = C_1^{(+)} = 0$ for the plus-mode 
and $A_1^{(-)} = A_2^{(-)} = C_2^{(+)} = 0$ for the minus-mode lasing to generate purely 
outgoing waves. Bimodal-lasing is provided once all four conditions in (\ref{twomodelasing}) 
are imposed. Notice that plus-mode lasing yields lasing in both sides of the slab due to the 
effective refractive index $\tilde{\fn}_{+}$ and a lasing from left-hand side due to 
$\tilde{\fn}_{-}$. Likewise, minus-mode lasing leads to a bidirectional lasing from 
$\tilde{\fn}_{-}$ and a left side lasing due to $\tilde{\fn}_{+}$. This reveals the 
distinctive character of TWS laser slab. Fig.~\ref{figplusminuslasing} clearly demonstrates 
these lasing behaviors of plus and minus-modes.

\section*{Lasing Conditions in Plus/Minus-Modes and Related Parameters}
\label{S3}

\begin{figure*}
    \begin{center}
    \includegraphics[scale=.55]{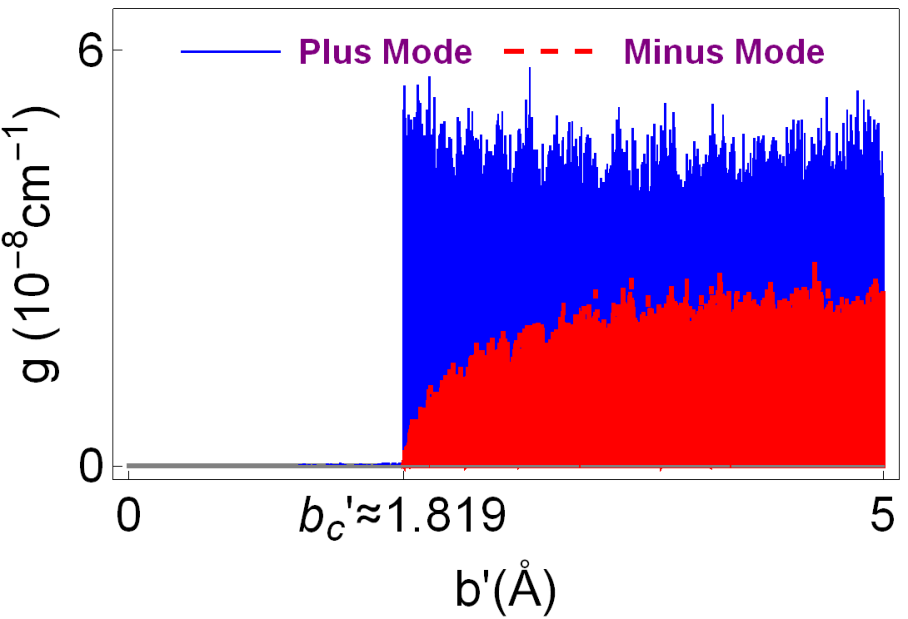}
    ~~\includegraphics[scale=.46]{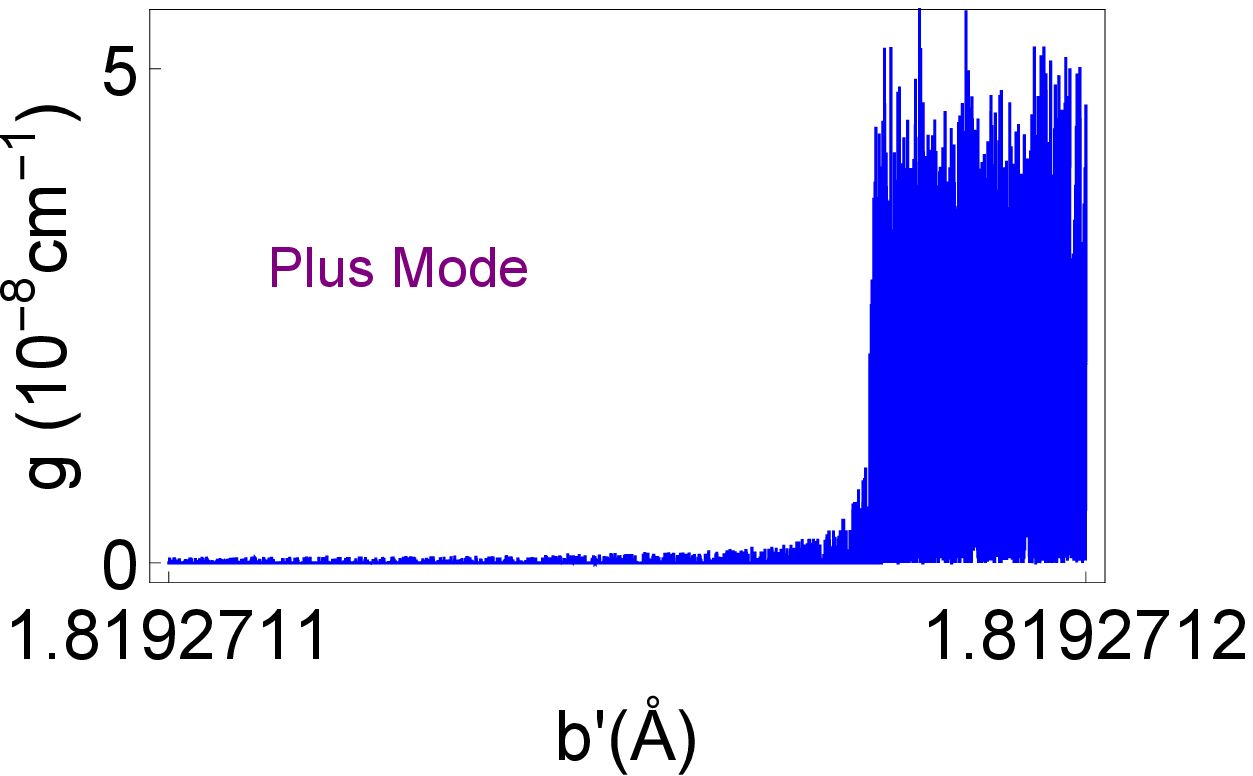}
    ~~\includegraphics[scale=.46]{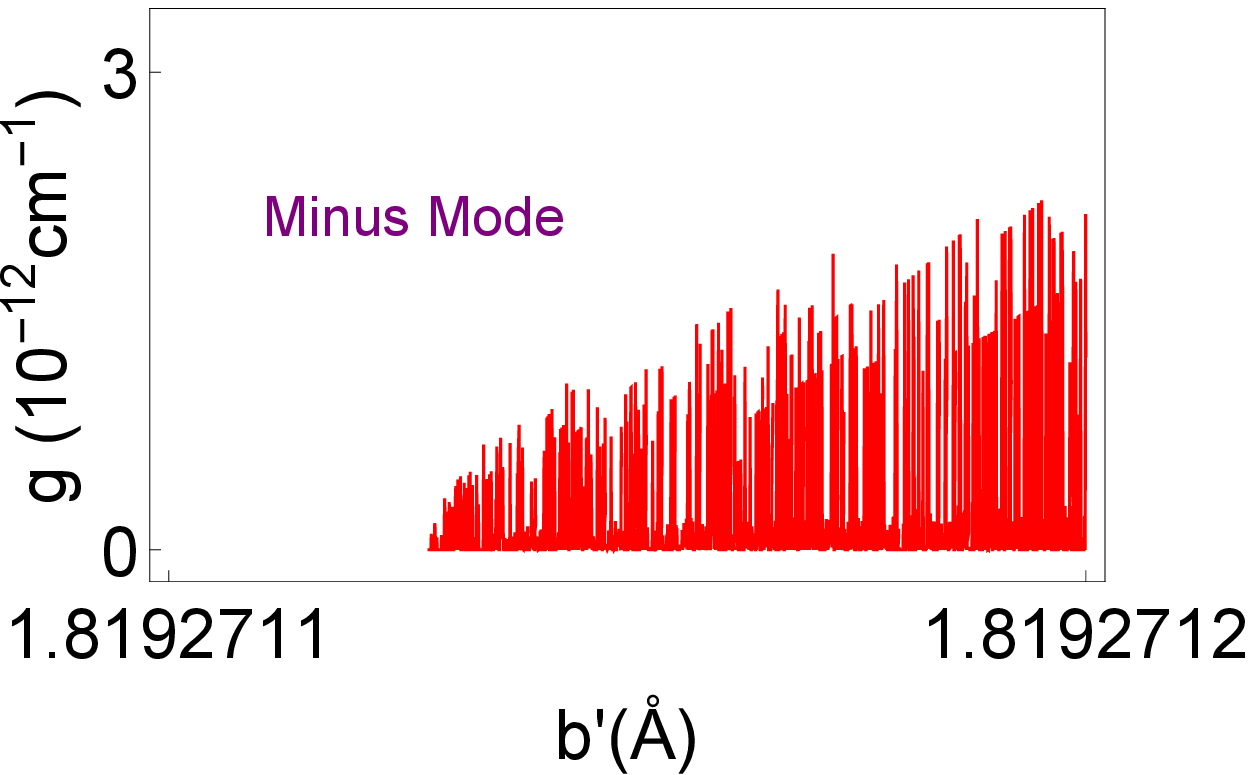}
    \caption{(Color online) Gain coefficient as a function of Weyl node separation $b'$ 
corresponding to the plus and minus modes for the configuration of the TWS slab system 
given in (\ref{specifications}). Here, $b'_c$ denotes the critical $b'$ value which 
corresponds to the minimum $b'$-value that allows lasing. Last two Figs. indicate the 
region that $b'_c$ takes effect. Horizontal gray line matches up to the zero gain limit.}
    \label{figbprime}
    \end{center}
    \end{figure*}

\begin{figure}
    \begin{center}
    \includegraphics[scale=.50]{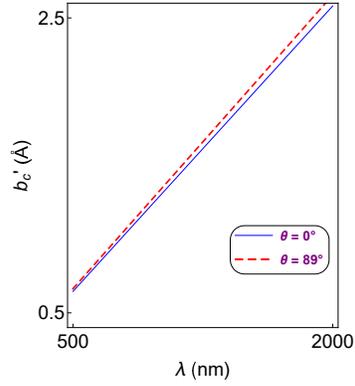}
    \caption{(Color online) Dependence of critical $b'_c$-value on the wavelength $\lambda$ 
for various incidence angles, corresponding to both plus and minus-mode lasing.}
    \label{figbprime2}
    \end{center}
    \end{figure}

Uni- or bimodal lasing behavior of TWS slab described by Eq.~\ref{twomodelasing} can be further analyzed to reveal conditions for the necessary parameters of the corresponding optical system. An immediate and simultaneous computation of two equations in (\ref{twomodelasing}) gives rise to the following expression for each individual mode
  \be
  e^{2i\fK_j} = -\mathcal{U}_j + \sqrt{\mathcal{U}_j^2 + 1}\label{onemodelasingcond}
  \ee
with the following identifications
  \begin{align}
  \mathcal{U}_j &:= \frac{\mathcal{X}_j\mathcal{Z}_j \pm \mathcal{Y}_j \sqrt{\mathcal{X}_j^2-\mathcal{Y}_j^2+\mathcal{Z}_j^2}}{\mathcal{Y}_j^2-\mathcal{Z}_j^2}\notag\\
  \mathcal{X}_j &:= \frac{1}{2}\left\{\mathcal{A}_{2j}^2-\mathcal{A}_{1j}^2 +\sigma_{j}^2\left[\mathcal{A}_{3j}^2-\mathcal{A}_{4j}^2-\mathcal{A}_{5j}^2\right]\right\}\notag\\
  \mathcal{Y}_j &:= \frac{1}{2}\left\{\mathcal{A}_{2j}^2+\mathcal{A}_{1j}^2 -\sigma_j^2\left[\mathcal{A}_{3j}^2+\mathcal{A}_{4j}^2\right]\right\}\notag\\
  \mathcal{Z}_j &:= \mathcal{A}_{1j}\mathcal{A}_{2j} +\sigma_j^2\mathcal{A}_{3j}\mathcal{A}_{4j}\notag\\
  \mathcal{A}_{1j} &:= \sigma_{j} -\sigma_{-j} -2,~~~~ \mathcal{A}_{2j} :=\frac{\tilde{\fn}_{j}^2 -\mu^2(\sigma_{j}-1)(\sigma_{-j}+1)}{\mu\tilde{\fn}_{j}}\notag\\
  \mathcal{A}_{3j} &:= \frac{\mu}{\tilde{\fn}_{-j}}\left[(\sigma_{-j}-1)(\sigma_+-\sigma_-+1-2j)+1\right]\notag\\
  \mathcal{A}_{4j} &:=\frac{\mu^2(\sigma_{-j}+1)\left[(\sigma_{-j}-1)\mathfrak{a}_{j}+1\right] -j(\sigma_{-j}-1)\tilde{\fn}_{j}^2}{\tilde{\fn}_-\tilde{\fn}_+},\notag\\
  \mathcal{A}_{5j} &:= \frac{\sqrt{2}\mu}{\tilde{\fn}_{-j}},~~~~~\mathfrak{a}_{j}:=1+j(\sigma_j-1),\notag
  \end{align}
where the subindex $j = +~\textrm{and}~-$ corresponds to the plus and minus modes. Uni- or 
bimodal lasing behavior of our TWS system is controlled by the expression 
(\ref{onemodelasingcond}). We realize that surface conductivities $\sigma_j$ associated with 
each mode are the manifestation of bimodal lasing attitude, which stems from the presence of 
Weyl nodes, and thus the Kerr and Faraday rotations. If the conditions of forming TWS are 
removed or disturbed considerably, the condition of uni- or bimodal lasing in 
Eq.~\ref{onemodelasingcond} is violated. Notice that (\ref{onemodelasingcond}) is a complex 
expression displaying the behavior of system parameters of our optical setup. The most 
appropriate system parameters should be chosen for the emergence of optimal impacts. Thus, 
it can be explored extensively by means of relevant physical quantities containing 
significant consequences by splitting into the real and imaginary parts. In this direction, 
refractive index of TWS described by Eq.~\ref{refractive} leads to the effective refractive 
indices which are bifurcated as follows
  \be
  \tilde{\fn}_j \approx \tilde{\eta}_j + i \tilde{\kappa}_j, \label{njtilde}
  \ee
with the identifications
\begin{flalign}
\tilde{\eta}_j := \sec\theta\sqrt{\eta^2 -\sin^2\theta +\frac{2j\alpha bL\cos\theta}{\pi\fK}},~~~~~~~\tilde{\kappa}_j := \frac{\eta\kappa\sec^2\theta}{\tilde{\eta}_j}.\notag
\end{flalign}
Notice that expression for $\tilde{\fn}_j$ in Eq.~\ref{njtilde} makes use of the condition 
that most of the materials including TWS satisfy
\be
\left|\kappa\right| \ll \eta-1 < \eta.\notag
\ee
Next we introduce the gain coefficient $g$ by
\be
g:=-2k\kappa = -\frac{4\pi\kappa}{\lambda} = -\frac{2\fK\kappa}{L\cos\theta}.\label{gaincoefficient}
\ee
Therefore, uni- or bimodal spectral singularity condition expressed by 
Eq.~\ref{onemodelasingcond} gives rise to the following expressions for the gain coefficient 
$g_j$ and effective wavenumber $\fK^j$~\footnote{Here we use superindex $j$ to avoid abuse of 
notation in Eq.~\ref{ktildej}. $\fK^j$ simply refers to the wavenumber corresponding to the 
mode $j$. }
\begin{flalign}
g_j&= \frac{\tilde{\eta}_j\cos\theta}{2\eta L} \ln\Bigl|\mathcal{V}_j\Bigr|,\label{gainplusminusmode}\\
\fK^j&=\frac{1}{2\tilde{\eta}_j}\sin^{-1}\Bigl\{\frac{-\textrm{Im}[\mathcal{U}_j] \pm \sqrt{\mathcal{W}_j} \sin\left(\frac{\Phi_j}{2}\right)}{\sqrt{\mathcal{V}_j}}\Bigr\},\label{kplusminus}
\end{flalign}
where we define the associated quantities
\begin{align}
\mathcal{V}_j &:= \mathcal{W}_j + \textrm{Re}[\mathcal{U}_j]^2 + \textrm{Im}[\mathcal{U}_j]^2 \mp 2\sqrt{\mathcal{W}_j}\left[\sin\left(\frac{\Phi_j}{2}\right)\textrm{Im}[\mathcal{U}_j] + \cos\left(\frac{\Phi_j}{2}\right)\textrm{Re}[\mathcal{U}_j]\right],\notag\\
\mathcal{W}_j &:= \sqrt{\textrm{Re}[\mathcal{U}_j]^4 + \left(\textrm{Im}[\mathcal{U}_j]^2 -1\right)^2 + 2\,\textrm{Re}[\mathcal{U}_j]^2 \left(\textrm{Im}[\mathcal{U}_j]^2 +1\right)},\notag\\
\Phi_j &:= \tan^{-1}\left\{\frac{2\,\textrm{Re}[\mathcal{U}_j]\,\textrm{Im}[\mathcal{U}_j]}{1+\textrm{Re}[\mathcal{U}_j]^2 -\textrm{Im}[\mathcal{U}_j]^2}\right\}, \notag
\end{align}
and $\textrm{Re}$ and $\textrm{Im}$ correspond to the real and imaginary part of the relevant 
quantity, respectively. We explore the most appropriate system parameters through the 
expressions of gain coefficient $g_j$ and $\fK^j$ in Eqs.~\ref{gainplusminusmode} and 
\ref{kplusminus} for the emergence of optimal impacts. Thus, a comprehensive analysis of the 
involvement of system parameters is required to observe the final outcome. For this purpose 
we exhibit the general behaviors of system parameters through the physically applicable gain 
coefficient g plots. But, notice that $g_{-}$-value in (\ref{gainplusminusmode}) does not 
allow lasing for some wavelength configurations due to the presence of the parameter 
$\tilde{\eta}_{-}$. Although it is not explicitly seen in the expression for $g_{+}$, the 
same restriction holds for $g_{+}$ as well. This in turn implies that the distance $b$ 
between the Weyl nodes in the Brillouin zone must satisfy
   \be
   b < \frac{\pi^2(\eta^2-\sin^2\theta)}{\alpha\lambda}.\label{boundb}
   \ee
In position space, this corresponds to
   \be
   b' > \frac{2\alpha\lambda}{\pi(\eta^2-\sin^2\theta)},\label{boundbposition}
   \ee
if we introduce $b := 2\pi/b'$. In (\ref{boundbposition}), the lower bound of $b'$ can be 
identified by $b'_c := 2\alpha\lambda/\pi(\eta^2-\sin^2\theta)$. Thus we realize that the 
lower bound of $b'$ depends on the incidence angle $\theta$ and wavelength $\lambda$. But 
$\theta$ has little effect compared to $\lambda$. Therefore, once the wavelength is 
increased, corresponding $b'_c$ value is reduced in order to observe a lasing in both plus 
and minus modes. The following graphs display the justification of our calculations which 
are based on the following specifications of a nonmagnetic ($\mu \approx 1$) TaAs TWS slab 
system~\cite{kriegel, buckeridge, ramshaw, dadsetani, silfvast}
\begin{align}
&\eta \approx 6, \qquad \lambda=1400\,{\rm nm}, \qquad L = 1~\textrm{cm},\notag\\
&~~~~~~\theta = 30^{\circ} \qquad \textrm{and} \qquad b' = 0.05~{\AA}.
\label{specifications}
\end{align}
Fig.~\ref{figbprime} reveals the effect of the Weyl node separation $b'$ in position space 
on the gain coefficient $g$ for the realization of lasing. Gain values are tightly packed. 
Notice that both plus and minus nodes give rise to a minimal bound $b'_c$, below which no 
lasing occurs. The best lasing impact with high quality factor is acquired by opting a TWS 
material whose Weyl node separation is around the critical $b'$-value. $b'_c$-value raises 
slightly by increasing the incidence angle $\theta$, and significantly by the wavelength 
rise. Fig.~\ref{figbprime2} shows how $b'_c$ varies with respect to wavelength $\lambda$ and 
incidence angle $\theta$.

\begin{figure*}
    \begin{center}
    \includegraphics[scale=.60]{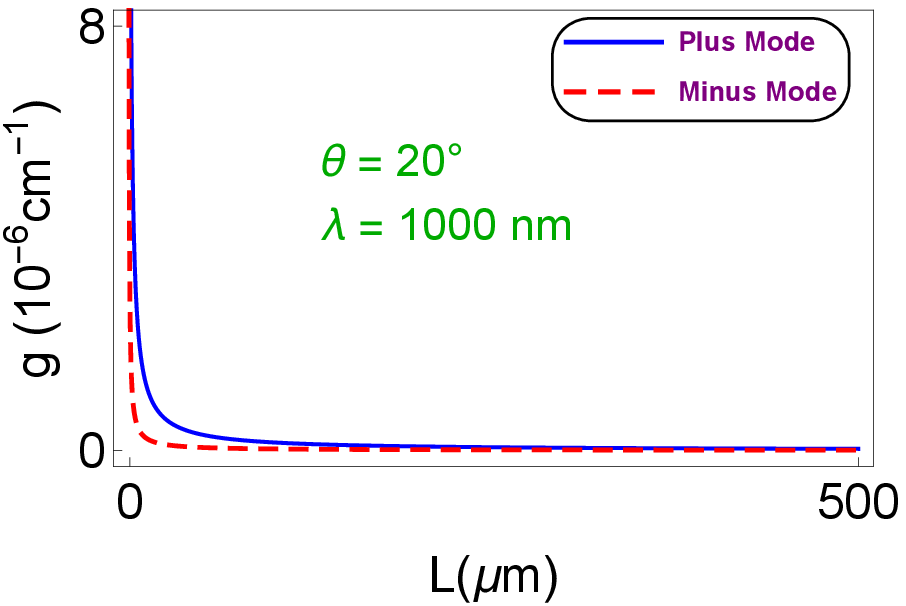}~
    \includegraphics[scale=.60]{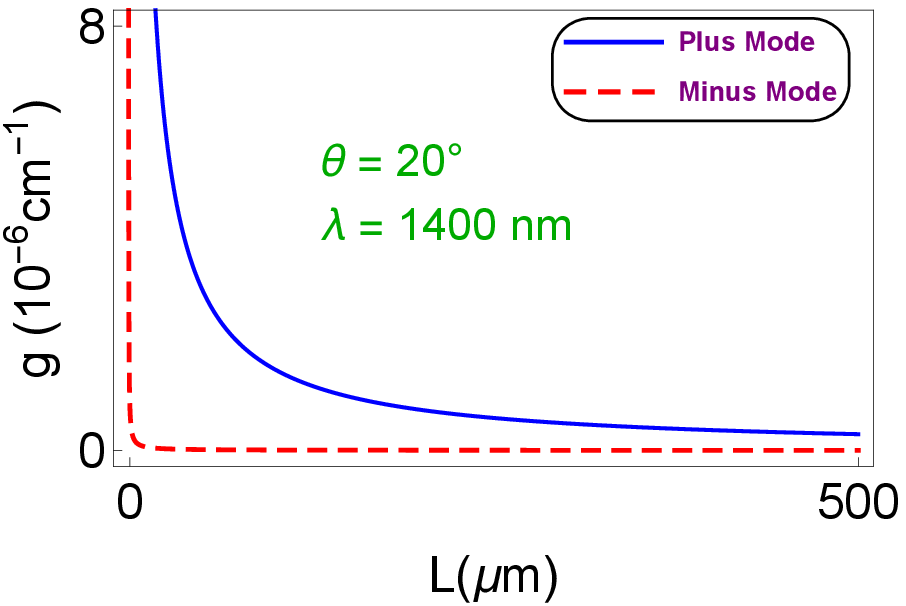}~
    \includegraphics[scale=.60]{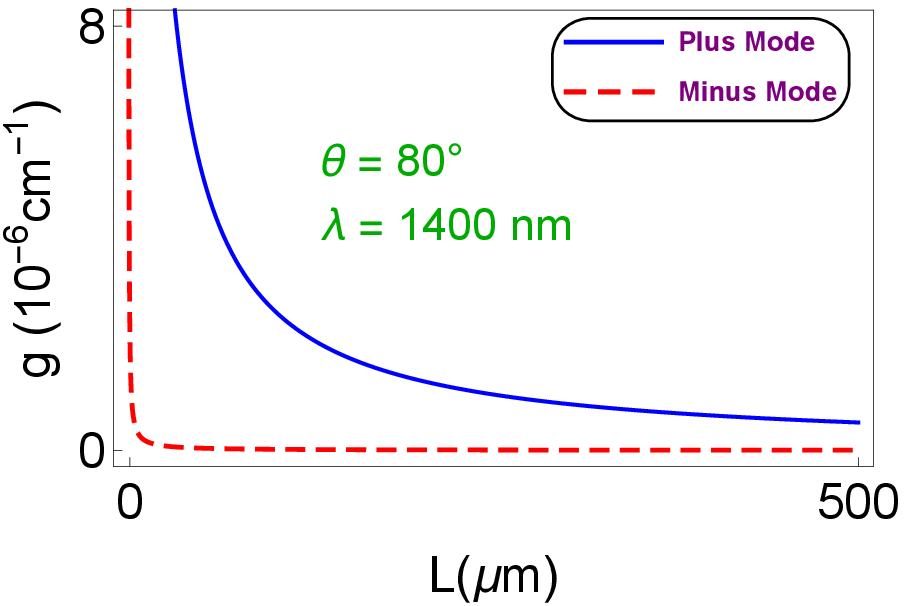}
    \caption{(Color online) Gain coefficient as a function of TWS slab thickness $L$ 
corresponding to various incidence angles and wavelengths. Increasing incidence angle and 
wavelength lead to higher gain values, and favor plus mode.}
    \label{figthickness}
    \end{center}
    \end{figure*}

Thickness of the TWS slab plays a rather distinctive role for lasing. The thicker the TWS 
slab is, the higher quality factor is achieved. This is obviously seen in 
Fig.~\ref{figthickness}. We realize that the less gain is obtained once we choose a 
relatively small wavelength $\lambda$ and incidence angle $\theta$. As we increase $\lambda$ 
and $\theta$, required gain values of plus and minus modes increase, indeed differ 
considerably.

\begin{figure*}
    \begin{center}
    \includegraphics[scale=.60]{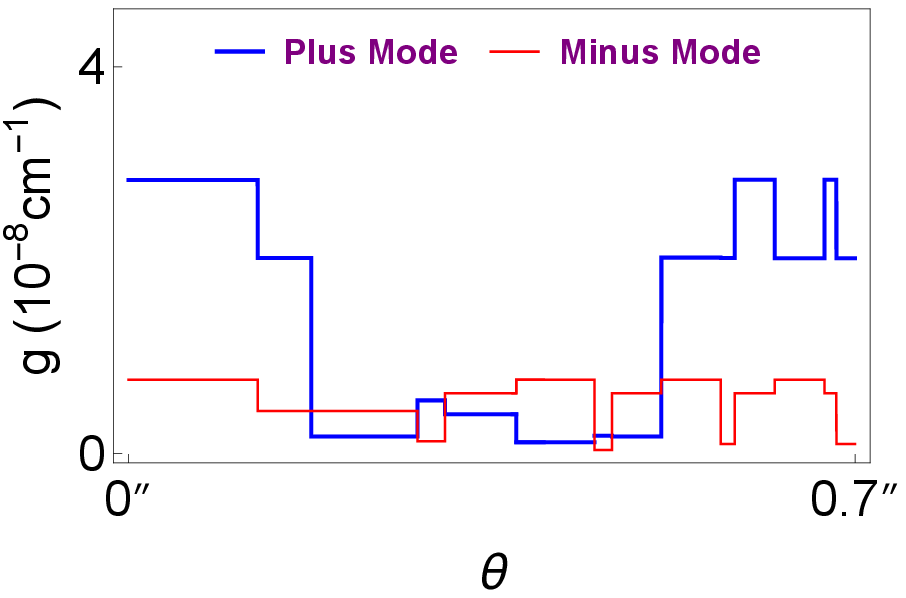}
    \includegraphics[scale=.60]{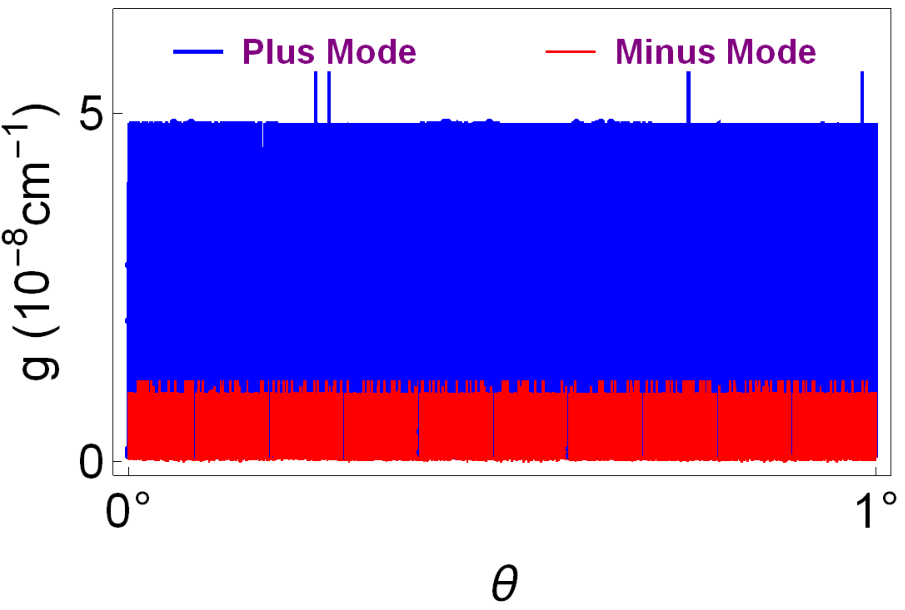}
    \includegraphics[scale=.60]{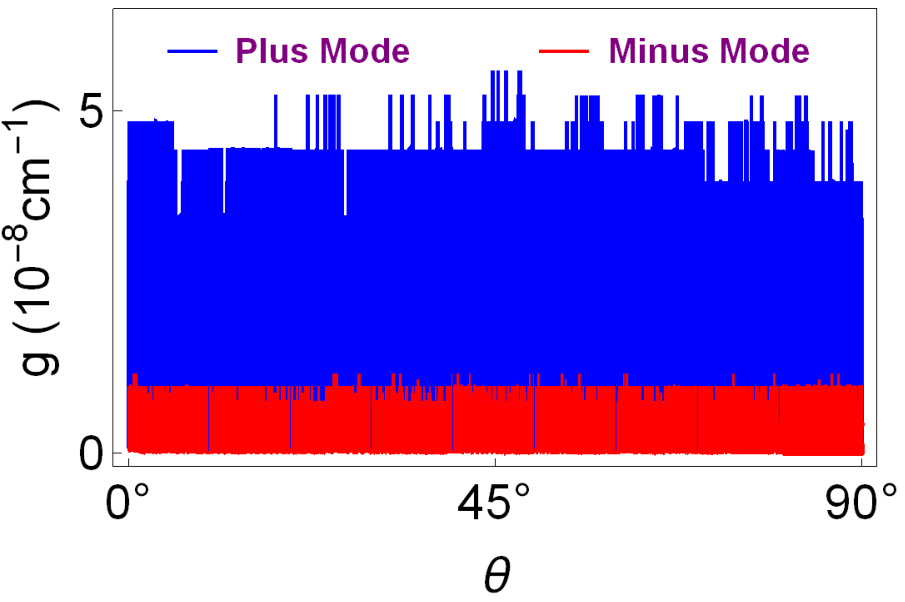}
    \caption{(Color online) Gain coefficient as a function of incident angle $\theta$. Here 
the symbol $^{\prime \prime}$ simply refers to the Arcsecond, which is given by 
$1^{\prime \prime} = 1^{\circ}/3600$. Notice that patterns are tightly packed once the range 
of the angle is increased.}
    \label{figangle}
    \end{center}
    \end{figure*}

\begin{figure}
    \begin{center}
    \includegraphics[scale=0.80]{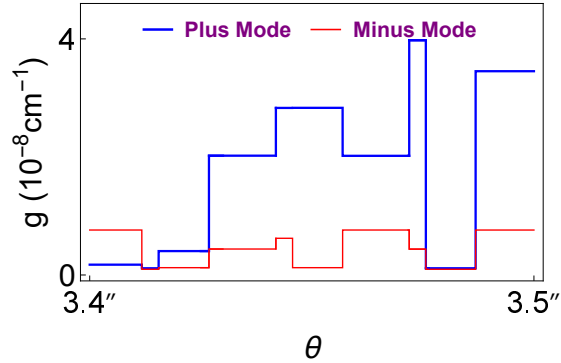}
    \caption{(Color online) Schematic diagram of the gain $g$ as a function of incident angle 
$\theta$ for the plus and minus lasing modes. Plus and minus modes overlap at angles 
$\theta \approx 3.42^{\prime \prime}$ and $\theta \approx 3.46^{\prime \prime}$ at the gain 
value $g \approx 1.17\times 10^{-9}~\textrm{cm}^{-1}$.}
    \label{p2m}
    \end{center}
    \end{figure}

Finally, Fig.~\ref{figangle} displays dependence of incidence angle on the gain coefficient 
$g$ to achieve lasing. We immediately realize the quantized patterns of gain such that only 
certain values are allowed for the lasing threshold condition. This quantization behavior 
is a clear indication of topological feature of our Weyl semimetal system. Also, notice 
the compact patterns that occur in both plus and minus modes. Again minus mode arises at 
small gain values compared to the plus mode. One can easily obtain the bimodal lasing using 
these patterns by observing the overlapping positions of plus and minus modes. For instance, 
Fig.~\ref{p2m} shows one such a case for a TWS system with parameters of incident angles 
$\theta \approx 3.42^{\prime \prime}$ and $\theta \approx 3.46^{\prime \prime}$ at gain 
value $g \approx 1.17\times 10^{-9}~\textrm{cm}^{-1}$.

\section*{Effect of Dispersion}
\label{S4}

\begin{figure*}
    \begin{center}
    \includegraphics[scale=0.80]{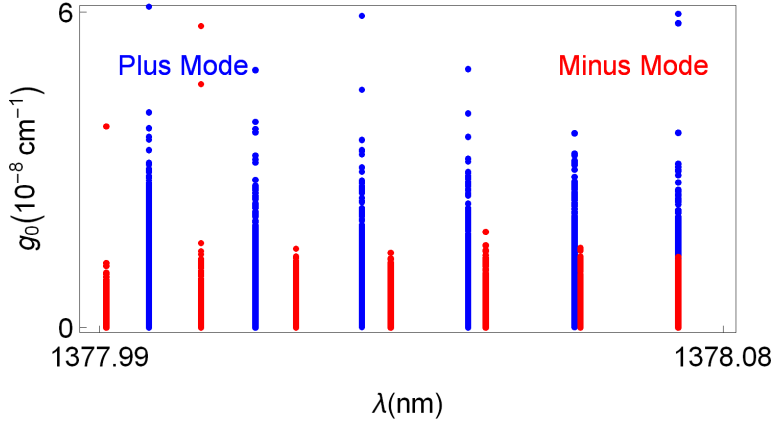}
    \caption{(Color online) The threshold gain $g_0$ as a function of wavelength $\lambda$ 
corresponding to both plus and minus modes. Plus mode is represented by blue dots while 
minus modes by red ones. Overlapping dots which take place at wavelength 
$\lambda\approx 1378.0735~\textrm{nm}$ imply the bimodal positions.}
    \label{p0}
    \end{center}
    \end{figure*}

If there exists a dispersion in the refractive index $\fn$, then we need to consider the 
effect of wavenumber $k$ on $\fn$. We imagine that active part of the TWS optical system 
with the gain ingredient is formed by doping a host medium of refractive index $n_0$ and 
its refractive index satisfies the dispersion relation
    \be
    \fn^2= n_0^2-
    \frac{\hat\omega_p^2}{\hat\omega^2-1+i\hat\gamma\,\hat\omega},
    \label{epsilon}
    \ee
where $\hat\omega:=\omega/\omega_0$, $\hat\gamma:=\gamma/\omega_0$, $\hat\omega_p:=\omega_p/\omega_0$, $\omega_0$ is the resonance frequency, $\gamma$ is the damping coefficient, and $\omega_p$ is the plasma frequency. $\hat\omega_p^2$ can be described in the leading order of the imaginary part $\kappa_0$ of $\fn$ at the resonance wavelength $\lambda_0:=2\pi c/\omega_0$ by the expression $\hat\omega_p^2=2n_0\hat\gamma\kappa_0$, where quadratic and higher order terms in $\kappa_0$ 
are ignored, \cite{pra-2011a}. By replacing this equation in (\ref{epsilon}), employing 
(\ref{refractive}), and neglecting the quadratic and higher order terms in $\kappa_0$, we 
obtain the real and imaginary parts of refractive index as follows \cite{CPA, pra-2011a}
     \begin{flalign}
    \eta\approx n_0+\frac{\kappa_0\hat\gamma(1-\hat\omega^2)}{(1-\hat\omega^2)^2+
    \hat\gamma^2\hat\omega^2},\kappa\approx\frac{\kappa_0\hat\gamma^2\hat\omega}{(1-\hat\omega^2)^2+
    \hat\gamma^2\hat\omega^2}\label{eqz301}
    \end{flalign}
$\kappa_0$ can be written as $\kappa_0=-\lambda_{0}g_0/4\pi$ at resonance wavelength 
$\lambda_0$. Substituting this relation in (\ref{eqz301}) and making use of 
(\ref{refractive}) and (\ref{onemodelasingcond}), we can determine the $\lambda$ and $g_{0}$ 
values for the uni- or bimodal TWS laser. In our optical configuration, TWS slab material 
holds the following values of the parameters refractive index, resonance wavelength, and 
corresponding $\hat\gamma$ value \cite{buckeridge, ramshaw, dadsetani, silfvast}:
    \begin{align}
    n_0=6, \qquad &\lambda_0=1378\,{\rm nm}, \qquad
    \hat\gamma=0.033,\notag\\
     \theta&=30^{\circ}, \qquad L=1~\textrm{cm}.
    \label{specifications2}
    \end{align}
Spectral singularity points which give rise to the lasing threshold conditions are explicitly 
calculated for various parameters in Table~\ref{table:config} for our setup of the TWS slab 
in (\ref{specifications2}). Fig.~\ref{p0} displays the positions of lasing points in 
$\lambda-g_0$ plane around the resonance wavelength, which yield divergent quality factors. 
Obviously, the wavelength layout corresponding to each particular mode is incompatible with 
the reciprocal mode unless they do coincide, in which case a bimodal lasing occurs. In our 
case, $\lambda\approx1378.0735~\textrm{nm}$ is the wavelength for the bimodal lasing. It is 
remarkable to realize that only particular discrete wavelengths allow lasing whereas 
threshold gain values are rather small compared to regular optical slab materials. This is 
encountered only with a TWS slab laser. If one considers different wavelength intervals, it 
is just enough to extend Fig. 9 to all wavelength ranges including ultraviolet, visible or 
infrared. But this time, corresponding gain values would be slightly higher since we move 
away from the resonance condition, see \cite{CPA, CPA-1, CPA-2, CPA-3}.
\begin{table}[ht]
\centering 
\begin{tabular}{c c c c} 
\hline\hline 
$\kappa$ & $g_0~(\textrm{cm}^{-1})$ & $\lambda~(\textrm{nm})$ & Type of Lasing \\ [0.5ex] 
\hline 
$-1.800\times 10^{-13}$ & $1.6419\times 10^{-8}$ & $1377.997$ & Plus Mode \\ 
$-6.341\times 10^{-14}$ & $5.7827\times 10^{-9}$ & $1378.0278$ & Plus Mode \\
$-1.539\times 10^{-13}$ & $1.4031\times 10^{-8}$ & $1378.0586$ & Plus Mode \\
$-4.089\times 10^{-14}$ & $3.7286\times 10^{-9}$ & $1377.9909$ & Minus Mode \\
$-1.238\times 10^{-13}$ & $1.1287\times 10^{-8}$ & $1378.0320$ & Minus Mode \\
$-1.927\times 10^{-14}$ & $1.7572\times 10^{-9}$ & $1378.0594$ & Minus Mode \\
$-1.237\times 10^{-13}$ & $1.1280\times 10^{-8}$ & $1378.0735$ & Bimodal \\
$-6.554\times 10^{-14}$ & $5.9773\times 10^{-9}$ & $1378.0735$ & Bimodal \\ [1ex] 
\hline 
\end{tabular}
\caption{ Types and corresponding configurations of a TWS slab laser with $L= 1~\textrm{cm}$ 
and $\theta = 30^{\circ}$.} 
\label{table:config} 
\end{table}

\section*{Kerr and Faraday Rotations in Lasing and CPA}
\label{S5}
We now turn our attention to the discussion of the Kerr and Faraday rotations which need to 
be understood to unveil the lasing behaviors and characteristics inside and outside the TWS 
slab, and explicitly see their impressive features. Since lasing is structured by the 
spectral singularities, we impose the spectral singularity conditions given in 
(\ref{twomodelasing}) or (\ref{onemodelasingcond}), and then impose the conditions 
$A_1^{(-)} = A_2^{(-)} = C_1^{(+)} = 0$ for plus-mode and 
$A_1^{(-)} = A_2^{(-)} = C_2^{(+)} = 0$ for minus-mode lasing to obtain necessary 
configurations of electric fields inside and outside the TWS slab, see~\cite{lastpaper} for 
details of this calculation. When we attain the spectrally singular fields, we can easily 
construct spectrally singular right and left circularly polarized fields 
$E_{\pm}(\mathbf{z})$, which are given by
\be
E_{+}(\mathbf{z}) = \sF_+(\mathbf{z})\,e^{i\fK \mathbf{x}\tan\theta}, \qquad E_{-}(\mathbf{z}) = \sG_+(\mathbf{z})\,e^{i\fK \mathbf{x}\tan\theta}. \notag
\ee
Hence, the Kerr and Faraday rotation angles are determined to be 
$\theta_K := \left(\arg E_{+}(\mathbf{z})-\arg E_{-}(\mathbf{z})\right)/2$ for 
$\mathbf{z}\not\in[0, 1]$ and 
$\theta_F := \left(\arg E_{+}(\mathbf{z})-\arg E_{-}(\mathbf{z})\right)/2$ for 
$\mathbf{z}\in[0, 1]$, respectively, where $\arg$ refers to the argument of corresponding 
quantity. Notice that when $\mathbf{z}<0$ of $\mathbf{z}>1$ we find the Kerr rotation angles 
$\theta_K$ which are equal in magnitude due to reciprocal response of the system. This 
observation leads to an important consequence about the CPA lasers built by TWS systems. 
The necessary and sufficient condition to generate a TWS CPA laser is to emit waves from 
both sides using polarizations exactly at Faraday angle $\theta_F$ with appropriate phases 
and magnitudes~\cite{lastpaper,CPA-8,CPA-9}. This is due to the main idea of a CPA that is 
time reversal symmetry of a regular laser~\cite{antilaser1, antilaser2, antilaser2-1, 
antilaser2-2, antilaser2-3, antilaser3, antilaser4, antilaser5}. Since this procedure 
requires special attention with excessive efforts, we reserve it for another study in its 
own right.

\section*{Concluding Remarks}
\label{S9}

In this study, we reveal basics of constructing a TWS laser by means of the transfer matrix 
approach. We employ the known features of TWS such that presence of the Weyl nodes gives 
rise to Fermi arcs configurations on relevant surfaces leading to magneto-optical effects. 
Alignment of the Weyl nodes determines how these magneto-optical effects response to the 
scattering of TWS slab. If the electromagnetic waves fall on the surface which has Fermi 
arcs, Maxwell equations turn out to be in simple familiar form, and do not contain any 
topological term. Besides, interesting case occurs when the electromagnetic waves are 
exposed to the surface with no Fermi arcs, which is what we have explored in the present 
paper. This case induces Maxwell equations with topological terms which give rise to 
coupled Helmholtz equations. We observe that this coupling refers to the inherent feature 
of Kerr and Faraday rotations of a generic TWS slab.

Our discussion strongly embraces the transfer matrix method such that the boundary conditions 
are highly effective~\cite{prl-2009, CPA, pra-2017a, cpa3}. Since the basic characteristics 
of a TWS slab manifest itself by the effects of the surfaces, this method becomes notable in 
this respect. We found out the transfer matrix, and obtained the condition of spectral 
singularities which lead to construct a TWS slab laser. We notice that our one dimensional 
problem turns into a two dimensional one by virtue of the Kerr and Faraday rotations. This 
gives a bimodal laser system corresponding to effective refractive indices 
$\tilde{\fn}_{\pm}$ in (\ref{birefrin}), which generate lasing in plus and minus modes as 
discussed in (\ref{onemodelasingcond}). These distinctive modes are peculiar to a TWS laser 
system, which can also be observed in any material with magneto-optical effects. We find out 
that exact lasing modes of a TWS slab are provided by the spectral singularity condition in 
(\ref{twomodelasing}). In particular, the plus mode lasing provides bidirectional lasing 
due to $\tilde{\fn}_{+}$ and left-side lasing due to $\tilde{\fn}_{-}$, whereas the minus 
mode lasing yields bidirectional one due to $\tilde{\fn}_{-}$ and left-side lasing due to 
$\tilde{\fn}_{+}$, see Fig.~\ref{figplusminuslasing}.

Lasing characteristics of these modes are obtained by the optimal control of system 
parameters. We have seen that the required gain value for a minus mode lasing is always 
less that the one for a plus mode. This gain value must be quantized, which is due to 
topological feature of TWS system. Also, relatively smaller TWS slab thickness is sufficient 
for the minus mode lasing. It is revealed that incident angles corresponding to uni- or 
bimodal lasing is so sensitive to tiny variations, and depends on adjusting definite gain 
value. For some distinct angles with appropriate gain values, one can obtain bimodal lasing 
as seen in Fig.~\ref{p2m}. This remarkable feature is explicitly seen in the case of 
dispersion effect in Fig.~\ref{p0} such that only some specific wavelengths give rise 
bimodal lasing. Once all other parameters are fixed, uni- or bimodal lasing occurs only at 
discrete wavelengths. Moreover, increasing wavelength gives rise to higher Weyl node 
separations as in Fig.~\ref{figbprime} when the remaining parameters are fixed. Considering 
all these behaviors of parameters, one can construct a highly effective TWS slab laser. We 
present some samples of these optimal parameters in Table~\ref{table:config}.

As a final consequence, this work pioneers the construction of a topological CPA laser. For 
the realization of a TWS CPA, one needs to adjust appropriate configuration of the whole 
topological optical setup because all incoming waves has to be absorbed inside the slab. 
For this to occur, we replace the gain by the loss, employ exact phase and magnitudes of 
the incoming waves, and finally set the polarization angle to Faraday rotation angle which 
can be computed with the help of the spectral singularity condition. Because of the 
arrangement of the Faraday rotation angle, a TWS CPA could be challenging. Our findings 
figure out this problem. We expect the results of this paper to guide experimental attempts 
for the realization of a concrete TWS laser and CPA. As a future perspective, it could be 
intriguing to discuss the dynamical cases, wherein the high-intensity lasers may damage the 
properties of the Weyl points. It is noted that the dynamical Floquet theory may be more 
applicable than the static considerations in this case.\\[6pt]


\appendix
\section*{Appendix}\label{boundarycond}

\subsection*{A. Modified Maxwell Equations}\label{modmax}

In the low energy limit of a TWS, spatially varying axion term plays a significant role in 
determining its electromagnetic response. The full action of corresponding TWS slab system is 
described by the sum of conventional and axionic terms as $S = S_0 + S_{\Theta}$
  \begin{align}
  S_0 &= \int \left\{-\frac{1}{4\mu_0}F_{\mu\nu}F^{\mu\nu} +\frac{1}{2}F_{\mu\nu} \mathcal{P}^{\mu\nu} - J^{\mu}A_{\mu}\right\} d^3x\,dt,\notag\\
  S_{\Theta} &= \frac{\alpha}{8\pi\mu_0}\int \left\{\Theta(\vec{r}, t)\varepsilon^{\mu\nu\alpha\beta}F_{\mu\nu}F_{\alpha\beta}\right\} d^3x\,dt,\notag
  \end{align}
where $F_{\mu\nu} = \partial_{\mu}A_{\nu}-\partial_{\nu}A_{\mu}$, $\mathcal{P}^{\mu\nu}$ tensor represents the electric polarization and magnetization respectively by $\mathcal{P}^{0i} = cP^{i}$ and $\mathcal{P}^{ij} =-\varepsilon^{ijk} M_{k}$. $J^{\mu}$ is the electric current four-vector and $\varepsilon^{\mu\nu\alpha\beta}$ is the totally antisymmetric tensor ($\varepsilon^{0123} =1$). Space and time dependent axion term is given by $\Theta(\vec{r}, t) = 2 \vec{b}\cdot\vec{r}-2b_0t$, where $\vec{b}$ and $b_0$ denote the separation of nodes in momentum and energy space respectively. In our case, $b_0$ is set to zero, since Weyl nodes are assumed to share the same chemical potential. If the action is varied with respect to $A_\mu$, the following equations of motion is obtained
  \be
  -\frac{1}{\mu_0}\partial_{\nu}F^{\mu\nu} + \partial_{\nu}\mathcal{P}^{\mu\nu} + \frac{\alpha}{2\pi\mu_0}\varepsilon^{\mu\nu\alpha\beta}\partial_{\nu}\left(\Theta F_{\alpha\beta}\right) = J^{\mu}
  \ee
It is obvious that expanding this equation yields the modified Maxwell equations given by 
Eqs.~\ref{equation1} and \ref{equation2} in the presence of the axion field term. For more 
details of the derivation of Maxwell equations in a TWS material environment, 
see \cite{Wilczek, Murchikova}.

\subsection*{B. Computation of Conductivities}\label{conduct}

To calculate the longitudinal $\sigma_{yy}$ and transverse $\sigma_{yx}$ conductivities of a 
TWS, we adopt the approaches given in \cite{conductiv1, weyl-conductivity}. We consider the 
simplest case with only two nodes located at $+\vec{b}$ and $-\vec{b}$, where 
$\vec{b} = b\,\widehat{e}_{z}$. Near the Weyl nodes, linearized Hamiltonian is given by
\be
H(\vec{k}) = \pm \hbar v_{F} \vec{\sigma}\cdot (\vec{k}\pm\vec{b}).\notag
\ee
where $v_{F}$ is the Fermi velocity, and $\vec{\sigma} = (\sigma_x, \sigma_y, \sigma_z)$ is 
the vector whose components are Pauli matrices. Therefore, conductivity 
$\sigma_{\alpha\beta}$ is obtained from Kubo formula as follows
\be
\sigma_{\alpha\beta} (\omega) = \frac{i}{\omega}\lim_{q\rightarrow 0}\Pi_{\alpha\beta}(q, \omega)\notag
\ee
In the absence of diamagnetic term, the polarization function $\Pi_{\alpha\beta}(q, \omega)$ 
is given by the current-current correlation function
\be
\Pi_{\alpha\beta}(q, i\omega_n) =  \frac{-1}{\mathcal{V}}\int_{0}^{\beta} d\tau\,e^{i\omega_n\tau} \braket{T_{\tau}\hat{\mathcal{J}}_{\alpha}(q, \tau)|\hat{\mathcal{J}}_{\beta}(-q, 0)}.\notag
\ee
where $\mathcal{V}$ is the volume of the system, and the current density operator 
$\hat{\mathcal{J}}$ is given by
\be
\hat{\mathcal{J}} = -\frac{\delta H}{\delta \vec{A}} = \pm e v_{F} \vec{\sigma}.\notag
\ee
Once we make analytic continuation $i\omega_n\rightarrow \omega + i0^{+}$, the real frequency 
behavior is obtained easily. Thus, for each node we obtain
\small
\begin{align*}
\Pi_{\alpha\beta}(\omega) = \frac{e^2 v_F^2}{\mathcal{V}}\sum_{i, i', \vec{k}}\frac{f(\varepsilon_{i'}(\vec{k}))-f(\varepsilon_{i}(\vec{k}))}{\hbar\omega +\varepsilon_{i'}(\vec{k})-\varepsilon_{i}(\vec{k})+i0^{+}} \braket{\vec{k}i|\sigma_{\alpha}|\vec{k}i'}\braket{\vec{k}i'|\sigma_{\beta}|\vec{k}i}
\end{align*}
\normalsize
where $f(x)=1 \Big/ (1+e^{\beta x})$ is the Fermi function. The expression 
$H(\vec{k})\ket{\vec{k}i} = \varepsilon_i (\vec{k})\ket{\vec{k}i}$ with $i=1, 2$ gives the 
quasiparticle energies and eigenstates. We can evaluate the longitudinal and transverse 
polarizations $\Pi_{\alpha\beta}(\omega)$ when the Fermi energy lies with nodes. Hence, in 
the low frequency limit, the longitudinal and transverse conductivities from both nodes are 
found to be expressions given in (\ref{conductivity1}) and (\ref{conductivity2}).

\subsection*{C. Boundary Conditions }

Boundary conditions across the surface $\mathcal{S}$ between two regions of space are given 
by the statements: 1) Tangential component of electric field $\vec{E}$ is continuous across 
the interface, $\hat{n}\times (\vec{E}_1-\vec{E}_2) = 0$; 2) Normal component of magnetic 
field vector $\vec{B}$ is continuous, $\hat{n}\cdot (\vec{B}_1-\vec{B}_2) = 0$; 3) Normal 
component of electric flux density vector $\vec{D}$ is discontinuous by an amount equal to 
the surface current density, $\hat{n}\cdot (\vec{D}_1-\vec{D}_2) =\rho^s$; 4) Tangential 
component of the field $\vec{H}$ is discontinuous by an amount equal to the surface current 
density, $\hat{n}\times (\vec{H}_1-\vec{H}_2) = \vec{\cJ}^s$. Here $\hat{n}$ represents the 
unit normal vector to the surface $\mathcal{S}$ from region 2 to region 1. In our optical 
configuration, we do not have the third condition since there is no normal component of 
the electric field. Thus, we obtain the boundary conditions as in Table~\ref{table2},
\begin{table*}[ht]
\centering
{
    \begin{tabular}{@{\extracolsep{4pt}}ccl}
    \toprule
     \\
    $\mathbf{z}=0$  & {~~~~~} &
    $\begin{aligned}
    &\sum_{j=1}^{2}\,\left[A_j^{(-)}+C_j^{(-)}\right] = \sum_{j=-}^{+}\,\left[B_1^{(j)}+B_2^{(j)}\right],\\[3pt]
    &\sum_{j=1}^{2}\,g_j\left[A_j^{(-)}+C_j^{(-)}\right] = \sum_{j=-}^{+}\,j\left[B_1^{(j)}+B_2^{(j)}\right],\\[3pt]
    &\mu\sum_{j=1}^{2}\,\left[\left(g_j + 2h_j\right)A_j^{(-)} - \left(g_j-2h_j\right)C_j^{(-)}\right] = \sum_{j=-}^{+}\,j\tilde{\fn}_j\left[B_1^{(j)}-B_2^{(j)}\right],\\[3pt]
    &\mu\sum_{j=1}^{2}\,\left[A_j^{(-)}-C_j^{(-)}\right] = \sum_{j=-}^{+}\,\tilde{\fn}_j\left[B_1^{(j)}-B_2^{(j)}\right]. \\[3pt]
    \end{aligned}$\\
    \hline
    &\\[-10pt]
    $\mathbf{z}=1$  & {~~~~~} &$\begin{aligned}
    &\sum_{j=1}^{2}\,\left[A_j^{(+)}e^{i{\fK}}+C_j^{(+)}e^{-i{\fK}}\right] = \sum_{j=-}^{+}\,\left[B_1^{(j)}e^{i{\fK}_j}+B_2^{(j)}e^{-i{\fK}_j}\right],\\[3pt]
    &\sum_{j=1}^{2}\,g_j\left[A_j^{(+)}e^{i{\fK}}+C_j^{(+)}e^{-i{\fK}}\right] = \sum_{j=-}^{+}\,j\left[B_1^{(j)}e^{i{\fK}_j}+B_2^{(j)}e^{-i{\fK}_j}\right],\\[3pt]
    &\mu\sum_{j=1}^{2}\,\left[\left(g_j+2h_j\right)A_j^{(+)}e^{i{\fK}}-\left(g_j-2h_j\right)C_j^{(+)}e^{-i{\fK}}\right] = \sum_{j=-}^{+}\,j\tilde{\fn}_j\left[B_1^{(j)}e^{i{\fK}_j}-B_2^{(j)}e^{-i{\fK}_j}\right],\\[3pt]
    &\mu\sum_{j=1}^{2}\,\left[A_j^{(+)}e^{i{\fK}}-C_j^{(+)}e^{-i{\fK}}\right] = \sum_{j=-}^{+}\,\tilde{\fn}_j\left[B_1^{(j)}e^{i{\fK}_j}-B_2^{(j)}e^{-i{\fK}_j}\right].
    \end{aligned}$\\[-8pt]
    &\\
    \hline
    \end{tabular}%
    }
    \caption{Boundary conditions for TE waves corresponding to TWS slab. Here effective 
indices $\tilde{\fn}_{\pm}$ are defined by (\ref{birefrin}), and functions $g_j$ and $h_j$ 
are given in (\ref{ghfunc}).}
    \label{table2}
\end{table*}
In this table, we used ${\fK} := k_z L$ and ${\fK}_j := {\fK}\,\tilde{\fn}_j$. $g_j$ and 
$h_j$ are defined as follows for convenience
    \begin{align}
    g_j &:=\left\{\begin{array}{cc}
    + & {\rm for}~~j = 1,\\
    - & {\rm for}~~j = 2
    \end{array}\right. \qquad
    h_j :=\left\{\begin{array}{cc}
    \sigma_{+} & {\rm for}~~j = 1,\\
    -\sigma_{-} & {\rm for}~~j = 2
    \end{array}\right. \label{ghfunc}
    \end{align}
where $\sigma_{\pm}$ is specified by Eq.~\ref{sigmapl}.

\section*{Acknowledgements}
We would like to thank referees for stimulating guidance which enhanced the quality of the 
paper. We acknowledge the support of Scientific Research Projects Coordination Unit of 
Istanbul University (IU BAP).

\section*{Author Contributions}
M. S. conceived the idea. M. S. wrote the original manuscript, M. S. and M. T. revised and edited the manuscript. G. O., M.
S. and M. T. contributed equally to the whole work including calculations and analysis.

\end{document}